\documentclass[10pt,letter,amsmath,amssymb,aps,prd,nofootinbib,notitlepage,superscriptaddress,twocolumn]{revtex4-2}
\usepackage[T1]{fontenc}
\usepackage[english]{babel}
\usepackage[utf8]{inputenc}
\usepackage{lmodern}
\usepackage{amsmath,amssymb,amsfonts,amsthm,amscd,mathrsfs}
\usepackage{amsthm}
\usepackage{comment}
\usepackage{geometry}
\usepackage[normalem]{ulem}
\usepackage{xcolor}
\usepackage{cancel}
\usepackage{hyperref}
\usepackage{placeins}
\usepackage{float}
\usepackage{mathtools}
\usepackage{soul}
\usepackage{graphicx}
\usepackage{tikz}
\usepackage{longtable}

\newcommand\ee{\end{equation}}
\newcommand\be{\begin{equation}}
\newcommand\eea{\end{eqnarray}}
\newcommand\bea{\begin{eqnarray}}
\newcommand{\bV}{\mathbf{V}}

\newcommand{\B}{\textrm{B}}
\newcommand{\F}{\textrm{F}}

\newcommand{\HH}{\mathcal{H}}

\newcommand{\beeq}{\begin{equation}}
\newcommand{\eneq}{\end{equation}}

\def\B{\mathrm{B}}
\def\F{\mathrm{F}}

\begin{document}

\title{Testing the equivalence principle across the Universe: \\ a model-independent approach with galaxy multi-tracing} 

\author{Sveva Castello}
\email{sveva.castello@unige.ch}
\affiliation{D\'epartement de Physique Th\'eorique and Center for Astroparticle Physics,
Universit\'e de Gen\`eve, Quai E. Ansermet 24, CH-1211 Gen\`eve 4, Switzerland}
\author{Ziyang Zheng}
\email{Zheng@thphys.uni-heidelberg.de}
\affiliation{Institut f\"ur Theoretische Physik, Universit\"at Heidelberg, Philosophenweg 16, 69120, Heidelberg, Germany}
\author{Camille Bonvin}
\email{camille.bonvin@unige.ch}
\affiliation{D\'epartement de Physique Th\'eorique and Center for Astroparticle Physics,
Universit\'e de Gen\`eve, Quai E. Ansermet 24, CH-1211 Gen\`eve 4, Switzerland}
\author{Luca Amendola}
\affiliation{Institut f\"ur Theoretische Physik, Universit\"at Heidelberg, Philosophenweg 16, 69120, Heidelberg, Germany}

\begin{abstract}
We present a test of the equivalence principle on cosmological scales. This cornerstone of general relativity has been tested with high precision for standard matter, but its validity for the unknown dark matter remains a crucial open question. We construct a measurable quantity $E_P$ that acts as a null test, i.e.~it deviates from unity whenever the weak equivalence principle is violated. This quantity can be directly measured from the cross-correlation of two different galaxy populations, thanks to the inclusion of large-scale relativistic corrections. A key feature of our approach is that it only involves minimal assumptions, without the need to specify the power spectrum shape, the background evolution, the growth rate of cosmic structure, the galaxy bias function or a model for the potential violation of the equivalence principle. We provide forecasts for the Dark Energy Spectroscopic Instrument and the Square Kilometre Array (SKA). While the relativistic corrections can be detected with high significance by both surveys, $E_P$ can be constrained up to an interesting level only by SKA, with a precision around 7--15$\%$ within the redshift range $z < 0.6$. 
\end{abstract}

\maketitle 

\begin{center}
{\bf Introduction}
\end{center}

The advent of increasingly accurate and extended cosmological data allows us to test fundamental aspects of physics in new regimes of space and time and for the unknown dark sector. In this work, we present a new model-independent approach to test a cornerstone of general relativity: the weak equivalence principle (EP), i.e.~the assumption that all bodies fall in the same way in a gravitational potential. While the EP has been confirmed with very high precision for Standard Model particles (see e.g.~\cite{Wagner:2012ui}), its validity beyond the Solar System and for the dark matter component remains unclear.\footnote{Only specific dark matter scenarios have been constrained up to galactic scales (see e.g.~\cite{Stubbs:1993}).} Our test addresses this key open question on cosmological scales, without the need to specify the mechanism responsible for a potential EP violation. This allows us to assess in a general way whether dark matter is subject to the same unscreened long-range gravitational force as baryons, and whether additional non-gravitational forces (so-called fifth forces) also impact its behavior.

Our approach relies on the large-scale distribution of galaxies, which provides a well-established laboratory to probe the fundamental properties of gravity (see e.g.~\cite{Moresco:2022phi,Amendola:2016saw}). However, standard galaxy clustering analyses usually focus on specific models beyond general relativity, for instance Horndeski scalar-tensor theories \cite{Horndeski:1974wa}, and often rely on known physics at early times and/or a standard background cosmological expansion. Moreover, cosmological tests of gravity generally assume that the EP holds (see e.g.~\cite{DESI:2024yrg}), and its validity for dark matter is rarely addressed. In this work, in contrast, we show that future galaxy surveys can test the EP in a highly model-independent way. 

We identify a key quantity that is directly measurable and acts as a null test of the EP, such that a deviation from unity in this quantity clearly signals a breaking of the EP, independently of any model. Three key features of our method allow us to reach this goal: 1) the inclusion of large-scale relativistic corrections; 2) the combination of different tracers of large-scale structure; 3) a parameterization that does not require any knowledge of the power spectrum shape, the background cosmological evolution, the growth rate of cosmic structure, and the galaxy bias.

Previous studies have already identified the first two ingredients \cite{Bonvin:2018ckp, Bonvin:2020cxp, Castello:2022uuu, Bonvin:2022tii, Castello:2024jmq}. However, these results rely on a fixed shape of the power spectrum at high redshift before the acceleration of the Universe started, based on the constraints from the Cosmic Microwave Background (CMB). In this work, we drop this assumption for the first time, allowing us to test a much wider range of theories. In particular, we can constrain models of gravity that do not recover general relativity at high redshift, but also dark matter models with additional interactions (with dark photons, self-interactions or interactions with early dark energy) that may modify the shape of the power spectrum at high redshift. Moreover, the method presented here does not rely on any known time dependence of the deviations from standard gravity and for the first time combines all ingredients in a directly measurable null test. 

\vspace{0.2cm}
\begin{center}
{\bf The Euler equation and the equivalence principle}
\end{center}

We adopt the perturbed Friedmann metric in the Newtonian gauge, $\mathrm ds^2 = a^2[-(1+2 \Psi)\mathrm d \tau^2 + (1- 2 \Phi)\mathrm d \mathbf{x}^2]$, with $\tau$ being the conformal time and $a$ the scale factor. With this setup, at late cosmological times the Universe can be described with four quantities encoding linear perturbations: the matter density fluctuation $\delta_m$, the matter velocity field $\mathbf{V}$ and the two gravitational potentials $\Phi$ and $\Psi$, denoting the spatial and temporal distortions in the spacetime geometry. The relations among these quantities are uniquely determined by the theory of gravity and the energy content of the Universe.

The EP is encoded in the Euler equation, relating the velocity $\mathbf{V}$ and the temporal distortion $\Psi$. 
In general, this equation in Fourier space takes the form \cite{Amendola:2003wa,Bonvin:2018ckp} 
\begin{align}\label{eq:euler_mod}
V'+ \big[1 + \Theta(k, z)\big] V -\frac{k}{\mathcal H} \big[1 + \Gamma(k, z)\big] \Psi= 0\,,
\end{align}
where a prime denotes a derivative with respect to $\ln a$, $\HH$ is the Hubble parameter in conformal time, and $V$ is the velocity potential in Fourier space, defined through $\mathbf{V}= i {\mathbf{k}} V/k$.
The quantities $\Theta$ and $\Gamma$ are free functions of scale ($k$) and redshift ($z$), which respectively encode a friction term and a fifth force acting on dark matter. Both vanish if the equivalence principle is respected. As shown in \cite{Bonvin:2018ckp}, this formulation covers a rich phenomenology, including scenarios where dark matter is non-minimally coupled to an additional scalar or vector field. As an example, the expressions for these functions in a coupled dark energy model can be found in \cite{Amendola:2003wa} and for Horndeski theories in~\cite{Castello:2023zjr}.

Our objective is to develop a test indicating whether the functions $\Theta$ and $\Gamma$ take a nonzero value in our Universe. In the spirit of being fully model-independent, we also allow for generic deviations from standard gravity encoded in a function $\mu_G$ in the Poisson equation,
\begin{equation}\label{eq: Poisson}
     \Psi =-\frac{3}{2}\mu_G (k,z)\left(\frac{\mathcal{H}}{k}\right)^2\Omega_m \delta_m\,,
\end{equation}
where $\Omega_m(z)$ is the matter density parameter at redshift $z$. Models where dark matter is subject to additional non-gravitational forces have $\mu_G=1$, whereas modified gravity theories with a different coupling to dark matter and baryons typically have $\mu_G\neq 1$.

As we shall see, the distribution of galaxies provides direct measurements of the quantity
\begin{equation}\label{eq:E_P}
E_P\equiv 1+\Theta-\frac{3\Omega_m\mu_G\Gamma}{2f} \, ,
\end{equation}
where $f \equiv \delta_m'/\delta_m$ is the growth rate of cosmic structure. The parameter $E_P$ provides a robust criterion since it deviates from unity whenever the Euler equation is not valid, regardless of the physical origin of this violation. More precisely, if gravity is not modified ($\mu_G=1$) but an additional non-gravitational force acts on dark matter ($\Theta, \Gamma \neq 0$), we have $E_P\neq 1$. If gravity couples differently to dark matter and baryons ($\Theta, \Gamma \neq 0$ and possibly $\mu_G\neq 1$), $E_P$ also deviates from unity. On the other hand, if gravity is modified in a way that respects the EP, then $E_P=1$. 

A physical interpretation of $E_P$ can be obtained considering a simple model with a constant dark matter-dark energy coupling $\tilde{\beta}$ and $\mu_G=1$ \cite{Amendola:1999er}. In this case, we have $E_P= 1+c_1\tilde{\beta}+c_2\tilde{\beta}^2$, with $c_{1,2}$ of order unity today. Therefore, $E_P$ deviates from unity proportionally to the strength of the EP-violating interaction. We also note that $E_P$ can be linked to the function $E^{\rm break}$ defined in Eq.~(4.1) in~\cite{Bonvin:2020cxp} by relating velocities and gravitational potential to the density field. In the following, we forecast constraints on $E_P$ from galaxy surveys. 

\vspace{0.2cm}
\begin{center}
{\bf A model-independent test from galaxy clustering}
\end{center}

Galaxy surveys measure the galaxy number count fluctuations, 
\begin{equation}
\Delta(\mathbf{\hat{n}}, z)\equiv [N(\mathbf{\hat{n}},z)-\bar N(z)]/\bar N(z)\, ,
\end{equation}
where $N$ is the number of galaxies per pixel detected in direction $\mathbf{\hat{n}}$ and at redshift $z$, and $\bar N$ denotes the average number per pixel. In the linear regime, the observable $\Delta$ is given by \cite{Bonvin:2011bg, Challinor:2011bk, Yoo:2009au}
\begin{equation}
\begin{split} \label{eq:Delta_galaxies}
\Delta(\mathbf{\hat{n}}, z)&=b_g\,\delta_m-\frac{1}{\HH}\partial_r(\bV\cdot\mathbf{\hat{n}}) \\
&+\frac{1}{\mathcal H}\partial_r\Psi+{\bV'}\cdot \mathbf{\hat{n}} +\alpha \mathbf V\cdot \mathbf{\hat{n}}\,,
\end{split}
\end{equation}
where $r$ is the comoving distance, $b_g$ is the galaxy bias and we have defined
\begin{equation}\label{eq:alpha}
    \alpha \equiv 1-5s+\frac{5s-2}{\mathcal H r}-\frac{{{\HH}'}}{\mathcal H}+f^{\rm evol}\,.
\end{equation}
Here, $s$ is the magnification bias, accounting for the fact that surveys are flux limited, and $f^{\rm evol}$ is the evolution bias, encoding the evolution of galaxies.

The first line of Eq.~\eqref{eq:Delta_galaxies} contains the density contribution and the well-known redshift-space distortions (RSD), accounting for the impact of the galaxy velocities on the redshift~\cite{Kaiser:1987qv}. These terms are widely used to test gravity in clustering analyses, see e.g.~\cite{eBOSS:2020yzd}. The terms in the second line of Eq.~\eqref{eq:Delta_galaxies} are relativistic corrections, suppressed in Fourier space by one power $\HH/k$ with respect to density and RSD. These corrections involve Doppler contributions proportional to $\bV$ and $\bV'$ and the gravitational redshift effect proportional to $\Psi$, which encodes the effect of the distortion of time inside a gravitational potential. We will see that the relativistic terms, in particular gravitational redshift, provide key information to test the EP.

The quantity $\Delta$ contains additional relativistic corrections suppressed by $\big(\HH/k\big)^2$. In the following, we choose to work in a regime where these are negligible, imposing cuts in $k$, to simplify the modeling of the signal. This does not remove crucial information since the terms in Eq.~\eqref{eq:Delta_galaxies} already contain $\Psi$ and $\mathbf{V}$, which are the only two relevant quantities to test the EP. Moreover, removing low-$k$ modes only leads to a marginal loss of constraining power, as these are strongly affected by cosmic variance.\footnote{$\Delta$ also contains some integrated contributions such as gravitational lensing, which were shown to be negligible in the low redshift regime ($z < 1$) relevant for this work \cite{Jelic-Cizmek:2020pkh, Euclid:2021rez, Tansella:2017rpi}.} 

As usual in this kind of analysis, we assume that the non-linear gravitational coupling between dark matter halos and their galaxy content does not generate a sizable velocity bias at linear order (even in presence of an EP violation), so that we equate the galaxy velocity with the dark matter velocity. We also neglect the contribution to the galaxy velocity due to the baryons (obeying the EP), as this was shown to have a negligible impact \cite{Castello:2022uuu}.

In the following, we work in Fourier space, where Eq.~\eqref{eq:Delta_galaxies} takes the form
\begin{align}
\label{eq:Delta_Fourier}
\Delta(\mathbf{k},\mathbf{\hat{n}},z)&=b_{g}\,\delta_m-\frac{\mu^{2}}{\lambda} V-i\frac{\mu}{\lambda} \Psi+i\mu V' +\alpha i\mu V\, ,
\end{align}
with $\lambda\equiv{\mathcal H}/k$ and $\mu\equiv \mathbf{\hat{k}}\cdot\mathbf{\hat{n}}$. Using the Euler equation~\eqref{eq:euler_mod} and the Poisson equation~\eqref{eq: Poisson}, together with the continuity equation $V = -\lambda f\delta_m$, we can rewrite $\Delta$ in terms of $\delta_m$. This yields
\begin{align}
    \Delta = \delta_m  b_g \left(1+\beta_g  \mu^2-i\beta_g \, \mu \, \lambda \, ( \alpha - E_P)
    \right)\, ,\label{eq:Delta_EP}
\end{align}
where $\beta_g =f/b_g$. This expression clearly shows that the sensitivity to $E_P$ specifically arises from the relativistic corrections. 

We assume that the bias and growth function do not depend on scale in the linear $k$-range we consider here. Similarly, although in principle $E_P$ can depend on scale and time, we only consider its time dependence. This is a good approximation for many models~\cite{Bonvin:2018ckp}, including modified gravity~\cite{Raveri:2021dbu}, but can be lifted if the data provide enough constraining power. 

In practice, we can extract information from $\Delta$ in Eq.~\eqref{eq:Delta_EP} by considering two-point correlations. Relativistic corrections break the symmetry of such correlations, generating anti-symmetric contributions. In order to measure these, it is necessary to correlate differently biased populations of galaxies ~\cite{McDonald:2009ud,Bonvin:2013ogt}, for example a bright (B) and a faint (F) one with number counts $\Delta_\B$ and $\Delta_\F$. We can thus compute all auto- and cross-power spectra $P_{\Delta_{\rm L}\Delta_{\rm M}}\equiv \langle\Delta_{\rm L} \Delta_{\rm M}^*\rangle $ (with $\rm{L, M}=\{B,F\}$), keeping terms only up to order $\lambda$:
\begin{align}\label{eq:spectra1}
P_{\Delta_{\F}\Delta_{\F}} & =  \left(1+\beta_\F\mu^2\right)^2\frac{\beta _\B}{\beta_\F}S_g^2 BP\,,\\
\label{eq:spectra2}
P_{\Delta_{\B}\Delta_{\B}} & =  \left(1+\beta_\B\mu^2\right)^2\frac{\beta_\F}{\beta_\B}S_g^2 BP\,,  \\
P_{\Delta_{\F}\Delta_{\B}} &=\big[(1+\beta_\F\mu^2)(1+\beta_\B\mu^2) \nonumber\\
&\label{eq:spectra3}
 +i\lambda \mu \left( \tau_1+\mu ^2 \tau _2  \right)]S_g^2 BP\,,
\end{align}
where $P$ is the power spectrum at $z=0$ and $B \equiv b_{\B}b_{\F}G(z)^2$, with the linear growth factor $G \equiv \delta_m(z)/\delta_m(0)$. The quantity $S_{g}(k,\mu,z)$ is a damping factor encoding corrections due to non-linear RSD and spectroscopic errors, specified in Eq.~\eqref{eq:FoG}. The parameters $\tau_1$ and $\tau_2$ appearing in the cross-spectrum are given by
\begin{align}
\tau_1 & = \beta_\B\alpha_\B-\beta_\F\alpha_\F-E_P(\beta_\B-\beta_\F)\,, \label{eq:tau1}\\
\tau_2 & = \beta_\B\beta_F(\alpha_\B-\alpha_\F)\,.
\label{eq:tau2}
\end{align}
The expression for $\tau_1$ clearly shows that $E_P$ can be measured only if $\beta_\B\ne\beta_\F$, hence requiring two differently biased populations of galaxies.
We also note that $\alpha_{\B,\F}$ only depends on measurable quantities, i.e.~background terms and $s_{\B,\F}\,, f^{\rm evol}_{\B,\F}$, which can be measured from the mean distribution of galaxies. Hence, the spectra in Eqs.~\eqref{eq:spectra1}-\eqref{eq:spectra3} are only functions of $B(z),\sigma_g(z), \beta_{\B,\F}(z),E_P(z)$ and $P(k)$, where $\sigma_g(z)$ encodes the non-linear velocity dispersion entering in $S_g$, see Eq.~\eqref{eq:FoG}.

As discussed in~\cite{Bonvin:2013ogt}, the correlations are affected by wide-angle corrections from RSD, due the fact that the lines of sight to two correlated galaxies are not parallel. These terms are of the same order as the relativistic corrections and hence cannot be neglected. We calculate them following \cite{Beutler:2020evf}, but using the middle-point configuration for the correlations, where $\mu$ is the cosine angle between the vector $\mathbf{\hat{k}}$ and the direction to the mean point in the separation between the galaxies. With this, we find
\begin{align}
P_{\Delta_\F\Delta_\B}^{\mathrm{wa}}&=-\frac{2}{5}\frac{1}{{\mathcal H}r}(\beta_{\F}-\beta_{\B})i\lambda BP\nonumber\\&\times\left(5\mu^{3}-\frac{5}{2}\mu(\mu^{2}-1)\frac{d\log P}{d\log k}\right) \, ,\label{eq:wa-3}
\end{align}
to be added to Eq.~\eqref{eq:spectra3}. This contribution scales with $\lambda$ and is therefore of the same order as the relativistic terms in the cross-spectra. Wide-angle corrections on the auto-spectra are instead of order $\lambda^2$, thus negligible with respect to density and RSD.

\vspace{0.2cm}
\begin{center}
{\bf Fisher analysis}
\end{center}

Following the methodology of~\cite{Quartin:2021dmr}, we build the data covariance matrix for each bin in wavenumber $k=|\mathbf{k}|$ and cosine angle $\mu$ as
\begin{equation}
C=\left(\begin{array}{cc}
P_{\Delta_\F\Delta_\F} +P_{{\rm sn},\Delta_\F}& P_{\Delta_{\F}\Delta_\B}\\
P_{\Delta_\B \Delta_\F} & P_{\Delta_\B\Delta_\B}+P_{{\rm sn},\Delta_\B}
\end{array}\right)
\end{equation}
Here, we have included shot noise terms of the form
$   P_{{\rm sn},\Delta_{\B,\F}}=n^{-1}_{\B,\F}$,
where $n_{\B,\F}$ are the galaxy number densities. Since 
\begin{equation}
   P_{\Delta_\F\Delta_\B}= \langle \Delta_\F\Delta_\B^* \rangle= \langle \Delta_\F^*\Delta_\B \rangle^*=P^*_{\Delta_\F\Delta_\B}\,,
\end{equation}
the covariance matrix is Hermitian. The Fisher Matrix (FM) for the $n$-th $k$ bin and the $m$-th $\mu$ bin is the trace of  products of commuting Hermitian matrices and is therefore real,
\begin{equation}
F^{nm}_{\alpha\beta}=\frac{1}{2}\frac{\partial C_{ij}}{\partial\theta_{\alpha}}C_{jp}^{-1}\frac{\partial C_{pq}}{\partial\theta_{\beta}}C_{qi}^{-1}\, ,
\end{equation}
where $\theta_\alpha$ is the parameter vector. The total FM,
summed over the $k,\mu$ bins, reads
\begin{equation}
    F_{\alpha\beta,\rm tot}=\frac{V}{8\pi^2}\,\sum_{n,m} k_{n}^{2}\Delta k_n\Delta\mu_m F^{nm}_{\alpha\beta}\, .
\end{equation}
We choose a constant $\Delta\mu_m=0.1$ and $\Delta k_n=0.01 \, h/$Mpc.  The spectra are multiplied by factors accounting for the Finger-of-God (FoG) effect and the spectroscopic errors (see e.g.~\cite{Koda:2013eya,Howlett:2017asw}):
\begin{equation}
    S_{g}(k,\mu,z) = \exp[-(k\mu\sigma_{{\rm z}})^{2}/2] \, \exp[-(k\mu\sigma_{g})^{2}/2]\,,\label{eq:FoG}
\end{equation}
where
$\sigma_{{\rm z}}=\sigma_{0}(1+z)H(z)^{-1}$. We take $\sigma_{0}=0.001$ for the spectroscopic errors and leave the damping strengths $\sigma_{g}$ as free parameters in each redshift bin. 

To convert redshifts and angles to Fourier wavevectors, one needs an arbitrary reference cosmology, which we denote with the subscript $r$. In any other cosmology, $k$ and $\mu$ are distorted by the Alcock-Paczyński (AP) effect,
so that $\mu=\mu_{r}h /\alpha_{\rm AP}$ and $k=\alpha_{\rm AP} k_{r}$, where~\cite{2000ApJ...528...30M} 
\begin{equation}
    \alpha_{\rm AP}\,=\,\frac{1}{d}\sqrt{\mu_{r}^{2}(h^{2}d^2-1)+1}\,.\label{eq:alphaAP}
\end{equation}
The distortion therefore depends on $h\equiv E/E_r$ and on $d\equiv L_A/L_{Ar}$, where $E(z)\equiv H(z)/H_0$ is the dimensionless Hubble function and $L_A(z)\equiv H_0 D_A(z)$ is the dimensionless comoving angular diameter distance. 
Since $d$ is degenerate with the power spectrum at the linear level, we will consider the combination $h_d\equiv h d$ as a parameter in our analysis. We also note that the product ${\mathcal H}r$ appearing in Eq.~\eqref{eq:wa-3} can be written in terms of our parameters as $E_rL_{Ar}h_d$.

To summarize, the set of free parameters in each $z$-bin is
$
\{h_d,d,B,\sigma_g,\beta_\B,\beta_\F,E_P\}.
$
In addition, we parametrize the power spectrum shape in several wavebands in the first redshift bin $z_1$ and evolve it with the free function $B(z)$. We take the $k$-range $k\in (0.01-0.12) \, h/$Mpc with intervals $\Delta k=0.01 \, h/$Mpc, for a total of twelve $P(k)$ waveband parameters and seven $z$-dependent parameters per $z$-bin. 

We adopt uniform infinite priors (i.e.~no prior in the Fisher formalism) for all parameters except $d$. For the latter, we adopt a 3\%  prior, since $d$ can already be determined to this level with current data through the Hubble diagram or with even better precision when averaged over a redshift bin (e.g.~with the Pantheon+ supernova catalog \cite{Brout:2022vxf}). We fix $B(z_1)$ to unity in the first bin, as this quantity is fully degenerate with $P(k,z_1)$. Hence, $B(z)$ encodes the ratio $b_\B b_\F G^2$ with respect to the first bin. 

The last term of the wide-angle correction in Eq.~\eqref{eq:wa-3} introduces a significant complication, as it contains a derivative and hence needs to be evaluated taking finite differences among $k$ bins. However, we find that the entire wide-angle correction is subdominant in the $k,z$ range we explore, and the derivative term itself only changes the constraints on $E_P$ by less than 4\%. For this reason, we fix the derivative term to the fiducial value. A similar problem arises with the derivative term  $\mathcal{H}'/{\mathcal H}$ in Eq.~\eqref{eq:alpha}. Here too we find that neglecting this term changes the main constraints by at most $5\%$ within each bin, such that we can safely fix $\mathcal{H}'/{\mathcal H}$ to its fiducial value. Note that this procedure can be iterated when working with data: after obtaining constraints on $h$ and $P(k)$, we can re-calculate $d P/dk$ and $\mathcal{H}'/{\mathcal H}$ at the best fit and redo the analysis with this new fiducial, until convergence is reached.

As discussed below Eq.~\eqref{eq:Delta_galaxies}, we apply scale cuts in each redshift bin based on the requirement that the $\lambda$ hierarchies are preserved. This means that the terms of order $\lambda^0$ must be much larger than those of order $\lambda^1$, to avoid interference from higher-order terms. In practice, we enforce this condition by only considering the bins for which the Fisher matrix is positive definite, which leads us to exclude a few low-$k$ bins at each redshift. These conditions ensure that our results remain both physically meaningful and numerically robust. 

\vspace{0.2cm}
\begin{center}
{\bf Survey specifications and fiducial values}
\end{center}

We separately perform the analysis for two surveys: SKA Phase 2 (SKA2) \cite{Bull:2015lja} and the DESI Bright Galaxy Sample (BGS) \cite{DESI:2016fyo} (see Appendix \ref{app:survey_specs} for more information). In both cases, we split the galaxy sample into a bright and a faint population with a redshift-dependent flux cut, chosen such that the two populations have the same number of galaxies in each redshift bin. 

For SKA2, we assume a baseline galaxy bias difference $\Delta b = 1$ between the two populations, based on measurements performed in BOSS \cite{Bonvin:2023jjq}. Thus, we choose as fiducial values for the biases $b_{\B,\F}(z) = b_g(z) \pm 0.5$, where $b_g(z)$ denotes the bias of the total galaxy population. Note that neither $b_g(z)$ nor $\Delta b$ are measurable quantities and that the two free parameters encoding the biases in our analysis are instead $\beta_\B$ and $\beta_\F$, which we vary around their fiducial values. Since $E_P$ in Eq.~\eqref{eq:tau1} scales with the bias difference, but there are currently no predictions for the expected value of this difference in SKA2, we adopt a set of toy models with $b_{\B,\F}(z) = b_g(z) \pm 0.3, 0.6$ to study how the constraints on $E_P$ change. For DESI, we take the measured values for $b_{\B,\F}(z)$ in DESI-like simulations from \cite{Bonvin:2023jjq} and compute the fiducial values for $\beta_\B$ and $\beta_\F$ accordingly. 
 
Both the magnification and the evolution bias will be directly measurable from the average number of galaxies once the data become available. For the purpose of our analysis, we fix the magnification bias $s_{\B,\F}(z)$ to the fiducial values presented in Appendix \ref{app:survey_specs} for each survey and the evolution bias $f^{\rm evol}$ to $0$, as this was shown to have a subdominant impact on the results \cite{Bonvin:2023jjq,Castello:2023zjr}. 
For the cosmological parameters $P(k),d, h_d,B$ and the growth rate $f$, we adopt $\Lambda$CDM values from the final \textit{Planck} data release \cite{Planck:2018vyg}, 
while we take a fiducial value of $4.24$ Mpc$/h$ for the velocity dispersion $\sigma_g$ in each redshift bin \cite{Howlett:2017asw}. Finally, we assume no EP violation as fiducial, i.e.~$E_P=1$.

\vspace{0.2cm}
\begin{center}
{\bf Results and conclusions}
\end{center}
\begin{table}[h]
\caption{\label{tab:DESI32} DESI relative errors for the baseline bias scheme.}
\centering
\begin{tabular}{cccccccc}
$z$ & $\beta_F$ & $\beta_B$ & $h_d$ &  $E_P$ & $\tau_1$ &  $B$  & $\sigma_g$  \\ \hline
 0.15 & 0.1 & 0.099 & 0.045 & 28 & 0.14 & - & 0.48 \\
 0.25 & 0.11 & 0.1 & 0.05 & 21 & 0.18 & 0.065 & 0.39 \\
 0.35 & 0.11 & 0.1 & 0.047 & 15 & 0.23 & 0.061 & 0.34 \\
 0.45 & 0.16 & 0.14 & 0.062 & 30 & 1.03 & 0.068 & 0.41 \\
\end{tabular}
\end{table}

\begin{table}[h]
\caption{\label{tab:SKA32} SKA2 relative errors for the baseline bias scheme.}
\centering
\begin{tabular}{cccccccc}
$z$ & $\beta_F$ & $\beta_B$ & $h_d$   & $E_P$ & $\tau_1$& $ B$ & $ \sigma_g $ \\
\hline
 0.25 & 0.0026 & 0.0013 & 0.0019 & 0.07 & 0.0076 & - & 0.14 \\
 0.35 & 0.0024 & 0.0014 & 0.002 & 0.087& 0.015 & 0.033 & 0.12 \\
 0.45 & 0.0024 & 0.0016 & 0.0023 &0.12 & 0.03 & 0.031 & 0.096 \\
 0.55 & 0.0026 & 0.0019 & 0.0027 & 0.16 & 0.054 & 0.03 & 0.084 \\
 0.65 & 0.003 & 0.0023 & 0.0033 & 0.2& 0.086 & 0.029 & 0.077 \\
 0.75 & 0.0035 & 0.0028 & 0.004 & 0.25& 0.13 & 0.029 & 0.073 \\
\end{tabular}
\end{table}

\begin{figure}[h]
   \centering
\includegraphics[width=0.43\textwidth]{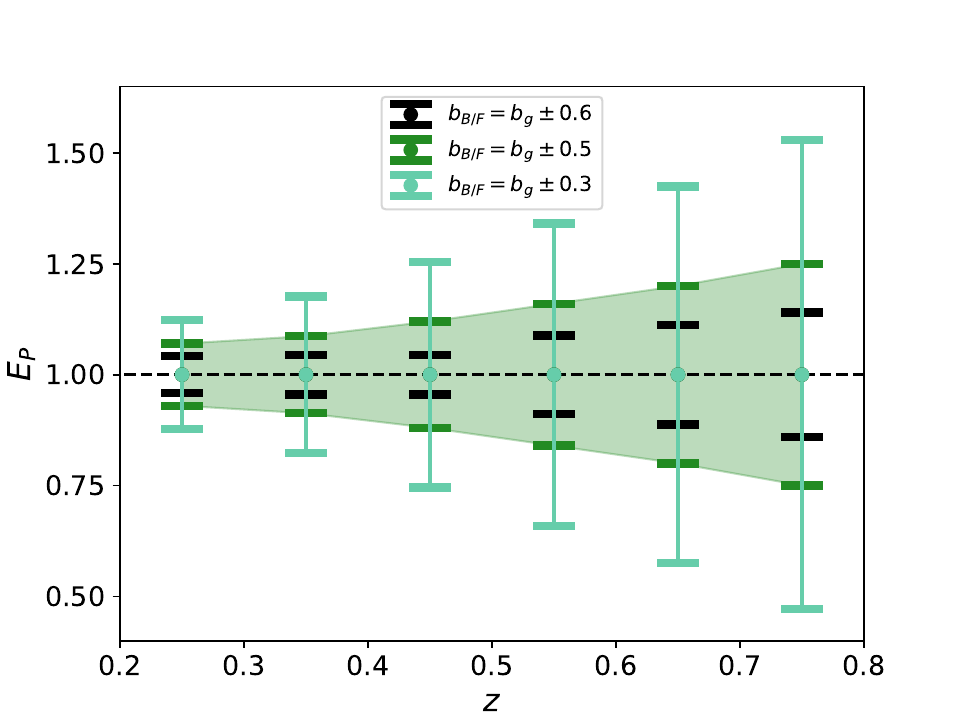}
    \caption{SKA2 1$\sigma$  relative error on $E_P$ as a function of redshift, for various bias schemes. The black dashed line represents the fiducial value $E_P=1$, while the green shaded region indicates the constraints for the baseline survey.} 
    \label{fig:E_P}
\end{figure}

In Table~\ref{tab:DESI32}, we show the relative uncertainties forecasted for DESI with the baseline bias scheme with $\Delta b = 1$. First, we assess whether the relativistic corrections can be detected at all. This can be achieved by estimating how well we can measure the $\tau_1$ coefficient in Eq.~\eqref{eq:spectra3}, which encodes the dominant relativistic contribution proportional to $\mu$ in the cross power spectrum $P_{\Delta_\F \Delta_\B}$. The quantity $\tau_1$ is not part of our set of parameters, but we can replace $E_P$ by $\tau_1$ using Eq.~\eqref{eq:tau1}. As we see from Table~\ref{tab:DESI32}, with DESI we obtain a constraint for $\tau_1$ of 14\% at $z = 0.15$, implying a signal-to-noise ratio of $1/0.14 \sim 7$. The constraints degrade for increasing redshift, but they are still robust, except in the last redshift bin. Hence, we conclude that relativistic corrections are expected to be well detected with DESI, in agreement with the results in \cite{Bonvin:2023jjq}. It is also possible to identify tailored cuts in magnitude to further boost the detection \cite{Montano:2024xrr}.

However, we find that DESI is unable to constrain $E_P$ to an acceptable level. Alternative ways of splitting galaxies, like density splits \cite{Paillas:2021oli} can in principle boost the bias difference, improving the constraints. For example, adopting a quite extreme bias difference of 2.0 around $b_g = (b_\B+b_\F)/2$ from the values in Table \ref{tab:DESI_specs}, we can constrain $E_P$ with an error of $\sim1.76$ at $z = 0.15$. 

For SKA2, we find that the relativistic corrections encoded in $\tau_1$ can be detected with a precision ranging between 0.8\%-13\%, see Table~\ref{tab:SKA32}. In this case, $E_P$ can be constrained to within $7-25\%$ in all redshift bins from 0.25 to 0.75, thanks to the larger volume, higher galaxy densities, and larger difference $\beta_\B-\beta_\F$ with respect to DESI. We include some plots of the full parameter space in Appendix \ref{app:plots}. 

To assess the robustness of the constraints, we have explored different configurations. Notably, we obtain that the relative error on $E_P$ changes by less than $3\%$ when adopting a lower $k_{\rm max}=0.1 \, h/$Mpc. Moreover, as expected from Eq.~\eqref{eq:tau1}, we find that the uncertainty on $E_P$ crucially depends on the bias scheme. In general, the best results are obtained by splitting the galaxy sample into populations with as much of a large $\Delta\beta$ as possible. For instance, adopting the alternative bias schemes $b_{\B,\F}=b_g\pm 0.3$ and $b_{\B,\F}=b_g\pm 0.6$, we obtain weaker or stronger constraints, respectively, as can be seen from Fig.~\ref{fig:E_P}. In addition, we find that the constraining power increases when the mean bias $b_g$ decreases. All our results are prior-independent, except for $d$, for which we adopted a data-motivated prior of 3\%. Even increasing the prior to 6$\%$, the final errors on $E_P$ for SKA2 do not change by more than a few percents. 

We find that the inclusion of the wide-angle correction only has a marginal impact on the constraints, except in the lowest bin, where it leads to an improvement of 54\%. This is due to the fact that the wide-angle correction depends on the combination $\Delta\beta$, which also appears in front of $E_P$. Therefore, improved constraints from the wide-angle terms help constraining $E_P$, also thanks to the $\mu$ terms carrying the AP effect and the further dependence on $P(k)$ in Eq.~\eqref{eq:wa-3}. 

In conclusion, SKA2 will provide promising and robust constraints on $E_P$. This demonstrates for the first time that it is possible to test the Equivalence Principle on cosmological scales independently of specific models for the mechanism leading to a potential violation, the power spectrum shape, the background cosmological expansion, the growth rate of structure, and the galaxy bias. This approach crucially relies on the measurement of large-scale relativistic corrections in the clustering of galaxies, and in particular on the term encoding the distortion of time in a gravitational potential. With the coming generation of galaxy surveys, our test paves the way to novel insights on the properties of dark matter and the laws of gravity on cosmological scales.

\section*{Acknowledgements}

We are grateful to Enea Di Dio for discussions on the wide-angle terms. SC and CB acknowledge support from the European Research Council (ERC) under the European Union's Horizon 2020 research and innovation program (grant agreement No.~863929; project title ``Testing the law of gravity with novel large-scale structure observables''). LA and ZZ acknowledge support from Deutsche Forschungsgemeinschaft (DFG, German Research Foundation) project 456622116 and from DFG Germany's Excellence Strategy EXC 2181/1 - 390900948 (the Heidelberg STRUCTURES Excellence Cluster). ZZ thanks the Heidelberg STRUCTURES Excellence Cluster for financially supporting her research visits to Geneva, and the University of Geneva for hospitality. SC is grateful for the hospitality at the Heidelberg University. The Mathematica code we employed to produce the Fisher matrix results is publicly available \cite{code}.

\newpage\onecolumngrid
\appendix

\section{Survey specifications}\label{app:survey_specs}

We present the specifications for the DESI and SKA2 surveys used in our forecasts and the resulting fiducial values for the parameters of the analysis. As specified in the main text, we choose $\Lambda$CDM with values from the final \textit{Planck} data release \cite{Planck:2018vyg}. Note that the observables do not depend on the value of the Hubble constant, which is just a dimensional factor in the units.

{\bf DESI Bright Galaxy Sample.}
The survey specifications are given in Table 2.5 in \cite{DESI:2016fyo}.  We set the galaxy bias and the magnification bias of the two galaxy populations to the values measured from DESI-like simulations in \cite{Bonvin:2023jjq} (Table 4, Case 1). In the lowest redshift bin, where no measurements could be performed, we adopt the same values as for $z = 0.25$, since we expect these functions to have a smooth redshift evolution. We again perform a flux cut requiring the same number of galaxies in each sample. We present all specifications in Table~\ref{tab:DESI_specs}.

{\bf SKA Phase 2.}
The survey specifications are given in Table 3 in \cite{Bull:2015lja}. The fiducial values for the galaxy bias were obtained with a fitting function for the SKA HI galaxy surveys based on \cite{Yahya:2014yva}. We compute the fiducial values of the magnification bias of the total galaxy population according to the fitting formula in Appendix A in \cite{Camera:2014bwa}, assuming a flux sensitivity limit of 5 $\mu$Jy. We then perform a flux cut as a function of redshift to split the total sample into two populations with the same number of galaxies. We present all specifications in Table \ref{tab:SKA2_specs}.

\begin{table*}[h]
\caption{\label{tab:DESI_specs}  DESI BGS survey specifications. Here and in the next table, the galaxy densities $n_g$ are multiplied by $10^3$ and expressed in units of $(h/{\rm Mpc})^3$ with $h=0.67$; the volume $V$ is in $({\rm Gpc}/h)^3$.} 
\centering
\begin{tabular}{ccccccccccccc}
$z$ & $V$ & $n_g $ & $b_{\rm B}$ & $b_{\rm F}$ &$s_\B$ &$s_\F$ & $\beta_\B$ & $\beta_\F$ &$\alpha_\B$ &$\alpha_\F$ & $\tau_1$ & $B$\\
\hline
 0.15 & 0.23 & 18.7 & 1.44 & 1.15 & 0.791 & -0.791 & 0.426 & 0.534 & 11.1 & -39.4 & 25.9 & 0.899 \\
 0.25 & 0.58 & 4.61 & 1.44 & 1.15 & 0.791 & -0.791 & 0.46 & 0.576 & 5.88 & -23.2 & 16.2 & 0.808 \\
 0.35 & 1.04 & 0.99 & 1.87 & 1.36 & 0.81 & -0.81 & 0.378 & 0.52 & 3.9 & -16.3 & 10.1 & 1.12 \\
 0.45 & 1.55 & 0.11 & 2.08 & 1.32 & 0.58 & -0.58 & 0.357 & 0.563 & 0.513 & -10.1 & 6.08 & 1.09 \\
\end{tabular}
\end{table*}

\begin{table*}[h]
\caption{\label{tab:SKA2_specs}  Baseline SKA2 survey specifications from \cite{Bull:2015lja}  as in the Aggressive case in \cite{Quartin:2021dmr}. }
\begin{tabular}{cccccccccccc}
$z$ & $V$ & $n_g $ & $b_g$ &$s_\B$ &$s_\F$ & $\beta_\B$ & $\beta_\F$   &$\alpha_\B$ &$\alpha_\F$ & $\tau_1$ & $B$\\
\hline
 0.25 & 1.2 & 121. & 0.674 & 0.371 & 0.0469 & 0.563 & 3.8 & -1.84 & -7.78 & 31.8 & 1. \\
 0.35 & 2.1 & 71.8 & 0.73 & 0.442 & 0.179 & 0.573 & 3.06 & -0.681 & -3.96 & 14.2 & 1.25 \\
 0.45 & 3.09 & 43.6 & 0.79 & 0.525 & 0.274 & 0.576 & 2.56 & 0.0138 & -2.29 & 7.9 & 1.49 \\
 0.55 & 4.11 & 26.8 & 0.854 & 0.607 & 0.357 & 0.573 & 2.19 & 0.379 & -1.36 & 4.8 & 1.72 \\
 0.65 & 5.11 & 17. & 0.922 & 0.682 & 0.44 & 0.565 & 1.9 & 0.533 & -0.779 & 3.1 & 1.95 \\
 0.75 & 6.06 & 10.9 & 0.996 & 0.755 & 0.52 & 0.554 & 1.67 & 0.576 & -0.428 & 2.2 & 2.19 \\
\end{tabular}
\end{table*}

\section{Additional plots}\label{app:plots}

\begin{figure*}[h]
    \centering
\includegraphics[width=0.75\textwidth]{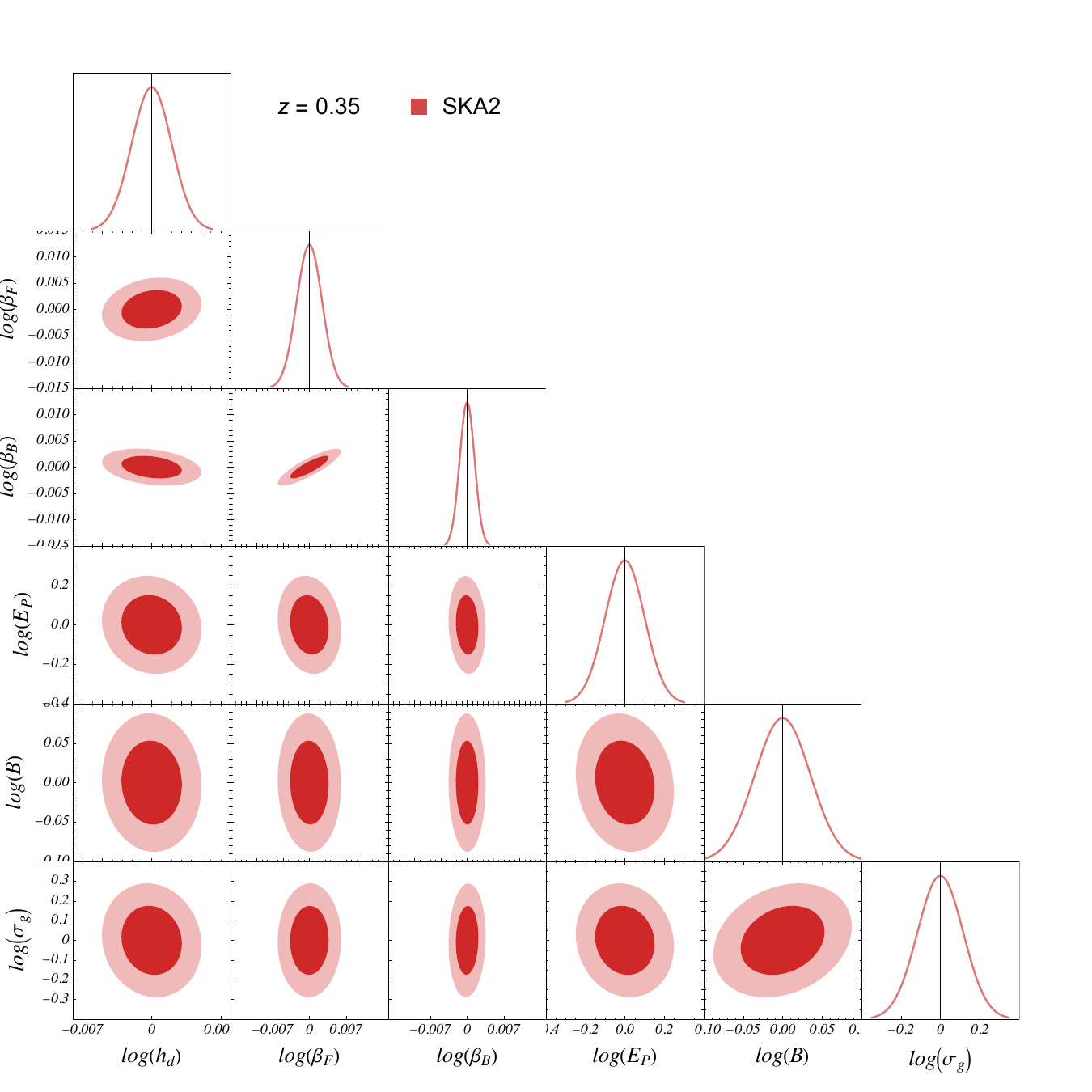}
    \caption{Corner plot for the baseline SKA2 survey at $z=0.35$ with one- and two-sigma regions. Since we always quote relative errors, the contour regions refer to the $\log$ of the parameters and, for each parameter $\theta$, are centered on $\theta/\theta_{\rm fid}$.}
    \label{fig:cp}
\end{figure*}

\begin{figure*}[h]
    \centering
\includegraphics[width=0.45\textwidth]{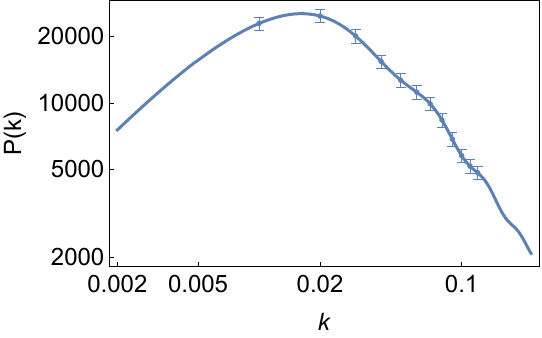}    \caption{Fiducial linear power spectrum at present time and error bars for the baseline SKA2 survey. Here, $k$ is in $h/$Mpc and $P(k)$ in $({\rm Mpc}/h)^3$.}
    \label{fig:pk}
\end{figure*}

For completeness, we also show in Fig.~\ref{fig:cp} the corner plot for all parameters at a representative redshift $z=0.35$, and in Fig.~\ref{fig:pk} the fiducial spectrum at $z=0$ with the forecasted errors. We find that the quantities $h_d,\beta_{\B,\F}$, which enter the spectra already at order $\lambda^0$, can be measured to within with SKA2, as can be seen from Table \ref{tab:SKA32}. The amplitude $B$ and the linear power spectrum, instead, can be measured within roughly $3-4\%$. 

\bibliographystyle{ieeetr} 
\clearpage
\bibliography{references_EP_paper}
\end{document}